\documentclass[twocolumn]{aastex631}

\newcommand{\kms}{km~s$^{-1}$\,}
\newcommand{\masyr}{mas~yr$^{-1}$\,}
\newcommand{\msun}{${\cal M}_\odot$\,}

\begin{document}

\renewcommand{\topfraction}{1.0}
\renewcommand{\bottomfraction}{1.0}
\renewcommand{\textfraction}{0.0}

\shorttitle{Speckle Interferometry at SOAR}
\shortauthors{Tokovinin et al.}

\title{Speckle Interferometry at SOAR in 2020 }

\author{Andrei Tokovinin}
\affil{Cerro Tololo Inter-American Observatory | NFSs NOIRLab Casilla 603, La Serena, Chile}
\email{andrei.tokovinin@noirlab.edu}

\author{Brian D. Mason}
\affil{U.S. Naval Observatory, 3450 Massachusetts Ave., Washington, DC, USA}
\email{brian.d.mason.civ@mail.mil}
\author{Rene A. Mendez}
\affil{Universidad de Chile,  Casilla 36-D, Santiago, Chile}
\email{rmendez@uchile.cl}
\author{Edgardo Costa}
\affil{Universidad de Chile,  Casilla 36-D, Santiago, Chile}
\author{Andrew W. Mann}
\affil{Department of Physics and Astronomy, University of North Carolina at Chapel Hill, Chapel Hill, NC 27599-3255, USA}
\author{Todd J. Henry}
\affil{RECONS Institute, Chambersburg, PA 17201, USA}

\begin{abstract}
The  results of  speckle  interferometric observations  at  the 4.1  m
Southern Astrophysical  Research Telescope (SOAR) in 2020,  as well as
earlier  unpublished data,  are given,  totaling 1735  measurements of
1288 resolved  pairs and non-resolutions of 1177  targets. We resolved
for  the first  time 59  new pairs  or subsystems  in  known binaries,
mostly among nearby  dwarf stars.  This work  continues our
long-term speckle program. Its main  goal is to monitor orbital motion
of  close binaries,  including members  of high-order  hierarchies and
Hipparcos pairs in the solar neighborhood. We also report observations
of  892 members  of young  moving  groups and  associations, where  we
resolved 103 new pairs.
\end{abstract} 

\section{Introduction}
\label{sec:intro}

This paper  continues the series  of double-star measurements  made at
the 4.1  m Southern Astrophysical  Research Telescope (SOAR)  with the
speckle   camera,   HRCam.    Previous   results  are   published   by
\citet[][hereafter            TMH10]{TMH10}           and           in
\citep{SAM09,Hrt2012a,Tok2012a,TMH14,TMH15,SAM15,SAM17,SAM18,SAM19}.
Observations  were  made during  2020,  but  this  work also  includes
earlier unpublished observations.

The structure  and content of this  paper are similar  to other paper of this series.
Section~\ref{sec:obs}   reviews  all  speckle   programs  that
contributed  to this  paper,  the observing  procedure,  and the  data
reduction.  The results are  presented in Section~\ref{sec:res} in the
form of  electronic tables archived  by the journal.  We  also discuss
new resolutions and new orbits  resulting from this data set.  A short
summary in Section~\ref{sec:sum} closes the paper.

\section{Observations}
\label{sec:obs}

\subsection{Observing programs}

As in previous years, HRCam (see Sect.~\ref{sec:inst}) was used during
2020  to   execute  several  observing  programs,   some  with  common
 targets.  Table~\ref{tab:programs}  gives an overview of
these programs  and indicates which observations are  published in the
present paper.  The numbers of observations  are approximate. Overall,
2348 observations were made during  2020.  Here is a brief description
of these programs.

{\it Orbits} of resolved  binaries. New measurements contribute to the
steady improvement of the quantity  and quality of orbits in the Sixth
Catalog of Visual Binary Star Orbits \citep{VB6}. See
\citet{Mendez2021} as an example of this work. 

{\it Hierarchical  systems} of stars  are of special  interest because
their architecture is relevant  to star formation, while dynamical
evolution   of  these   hierarchies  increases   chances   of  stellar
interactions  and  mergers  \citep{Toonen2020}.  We  followed  orbital
motions  of several  triple  systems  and used  these  data for  orbit
determinations \citep{TL2020,Tok2020,Tok2021a}.  Some observations made
in 2020 are published in the above papers. They are duplicated here to
provide a complete and homogeneous record of the SOAR speckle data.

{\it  Hipparcos  binaries} within  200\,pc  are  monitored to  measure
masses  of stars and  to test  stellar models,  as outlined  by, e.g.,
\citet{Horch2015,Horch2017,Horch2019}.   The  southern  part  of  this
sample is addressed at SOAR \citep{Mendez2017}.  This program overlaps
with the general work on visual orbits.

\begin{deluxetable}{ l l l l  } 
\tabletypesize{\scriptsize}    
\tablecaption{Observing programs
\label{tab:programs} }                    
\tablewidth{0pt}     
\tablehead{ \colhead{Program}  &
\colhead{PI}  &  
\colhead{$N$} & 
\colhead{Publ.\tablenotemark{a}} 
}
\startdata
Orbits                & Mason, Tokovinin    & 562 & Yes \\
Hierarchical systems  & Tokovinin           &  82 & Yes \\
Hipparcos binaries    & Mendez, Horch       &  267 & Yes \\
Neglected binaries    & R.~Gould, Tokovinin & 152 & Yes \\
Nearby K dwarfs     & T.~Henry              & 228  & Yes \\
Nearby M dwarfs     & E. Vrijmoet           &  354 & No \\
TESS follow-up        & C. Ziegler           & 355  & No \\
Young moving groups   & A. Mann              & 985  & Yes \\ 
Stars with RV trends  & B. Pantoja           & 195  & Yes
\enddata
\tablenotetext{a}{This columns indicates whether the results are
  published here (Yes) or deferred to future
  papers (No). }
\end{deluxetable}

{\it Neglected close binaries}   from  the Washington
Double Star Catalog, WDS \citep{WDS}, were observed as a `filler'
at low priority. In some cases, we resolved new inner subsystems, thus
converting classical visual pairs into hierarchical triples. Some
WDS pairs are moving fast near periastron, allowing calculation of
their first orbits after several observations at SOAR. 

{\it Nearby K and M dwarfs} were observed since 2018 on the initiative
of T.~Henry and E.~Vrijmoet. The goal is to assemble statistical data
on orbital elements, focusing on short periods. The sample includes
known and suspected binaries detected by astrometric monitoring,
Gaia, etc. Data on M dwarfs are being published by Vrijmoet et al. (2021, in
preparation), while observations of K dwarfs are reported here. 

{\it TESS  follow-up} received a substantial time  allocation in 2020,
continuing the program of  2018--2019. Its first results are published
by \citet{TESS}, the  paper with additional data is submitted
\citep{TESS2}. All  speckle observations  of TESS targets  of interest
are    promptly   posted   on    the   \href{https://exofop.ipac.caltech.edu/tess/}{EXOFOP web site.} These  data are used in
the growing number  of papers on TESS exoplanets,  mostly as limits on
potential companions to exohosts.

{\it Young moving groups} and  associations were selected as part of a
program aimed at characterizing planets and young stellar associations
with TESS \citep[the  THYME survey,][]{Newton2019}. These sources were
selected from because  they have been observed by  TESS and are likely
to be young or members  of nearby stellar associations reported in the
literature. The  majority of  stars (549) were  drawn from  the BANYAN
survey of  young moving groups within  150\,pc \citep{BanyanSigma}, as
well as 82 members  of the Scorpius–Centaurus (Sco-Cen) OB association
(Sco-Cen) from \citet{Rizzuto2015} excluding those already surveyed by
\citet{Sco} and  261 suspected pre-main-sequence  stars within 500\,pc
from \citet{Zari2018}. For both BANYAN and Sco-Cen members, membership
for  targets  were  determined  primarily  using  Bayesian  membership
probabilities  based on kinematics  of each  star and  the association
using Gaia DR2 astrometry. Sources from \citet{Zari2018} were selected
based on  their elevated position relative  to the main  sequence on a
color-magnitude diagram.  Because all  target selection relied on Gaia
DR2,  any  systematics present  in  the  Gaia  catalog (e.g.,  missing
binaries) will be present in the targets surveyed here. The names of these
objects in  the data tables begin  by `T' followed by  their number in
the TESS input catalog \citep[TIC,][]{Stassun2019}.

{\it Stars with  radial velocity trends} were monitored  since 2016 on
request from B.~Pantoja, with  the aim to resolve potential companions
causing these  trends \citep[e.g.][]{Pantoja2018}. Five  new pairs and
one triple (GJ 3260)  were  resolved at SOAR  and measured   during five
years.

If observations  of a given  star were requested by  several programs,
they are  published here even when the other program still  continues.  We
also publish here the measurements of  previously known pairs  resolved during
surveys, for example in the TESS follow-up.

\subsection{Instrument and Observing Procedure}
\label{sec:inst}

The   observations  reported   here  were   obtained  with   the  {\it
  high-resolution camera} (HRCam) -- a fast imager designed to work at
the 4.1  m SOAR telescope \citep{HRCAM}. The  instrument and observing
procedure  are  described  in  the  previous papers  of  these  series
\citep[e.g.][]{SAM19}, so only the  basic facts are reminded here.  We
used  mostly  the  near-infrared  $I$  filter  (824/170\,nm)  and  the
Str\"omgren $y$ filter (543/22\,nm), with a few observations made in
the $B$, $V$, and $R$ filters;  the transmission curves of HRCam
filters    are    given    in   the     \href{
http://www.ctio.noao.edu/soar/sites/default/files/SAM/\-archive/hrcaminst.pdf}{instrument manual}
In the standard observing mode, two series of 400 200$\times$200 pixel
images (image  cubes) are recorded.  The pixel  scale is 0\farcs01575,
hence the  field of  view is 3\farcs15;  the exposure time  is normally
24\,ms. For  survey programs  such as TESS  follow-up, we use  the $I$
filter and a 2$\times$2 binning,  doubling the field.  Pairs wider
than $\sim$1\farcs4  are observed  with a 400$\times$400  pixel field,
and the  widest pairs  are sometimes recorded  with the full  field of
1024 pixels (16\arcsec) and a 2$\times$2 binning.

The  speckle power  spectra are  calculated and  displayed immediately
after    the    acquisition    for    quick    evaluation    of    the
results. Observations  of close pairs are  accompanied by observations
of single  (reference) stars to  account for such  instrumental effects
as telescope  vibration  or aberrations.   Bright  stars can  be
resolved and measured below the  formal diffraction limit by fitting a
model to the  power spectrum and using the  reference.  The resolution
and contrast  limits of  HRCam are further  discussed in TMH10  and in
the previous papers of this series. 

A custom software helps to optimize observations by selecting targets,
pointing the telescope, and  logging. Typically, about 300 targets are
covered on  a clear night. The  observing programs are  executed in an
optimized  way, depending on  the target visibility,  atmospheric conditions,
and priorities, while minimizing  the telescope slews. Reference stars
and  calibrator binaries are  observed alongside  the main  targets as
needed.

During 2020, the SOAR telescope was  closed from March 18 to October 7
due  to COVID-19  pandemic.  The  number of  observations  obtained in
2020, 2348,  is less  than in 2018  and 2019.  The  sporadic telescope
vibration  that  affected  HRCam  observations  previously  \citep[see
  \S~3.5 in][]{HRCAM} was much less frequent in 2020.

\subsection{Data Processing and Calibration}
\label{sec:dat}

The data processing  is described in TMH10 and  \citet{HRCAM}.  We use
the standard speckle interferometry technique based on the calculation
of the power spectrum and the speckle auto-correlation function (ACF).
Companions  are detected  as  secondary  peaks in  the  ACF and/or  as
fringes in  the power spectrum.   Parameters of the binary  and triple
stars  (separation  $\rho$,  position  angle $\theta$,  and  magnitude
difference  $\Delta  m$)  are  determined by  modeling  (fitting)  the
observed  power  spectrum.   The  true  quadrant  is  found  from  the
shift-and-add (SAA)  images whenever possible  because the  standard speckle
interferometry determines position angles modulo 180\degr.

The pixel scale  and angular offset are inferred  from observations of
several  relatively  wide  (from  0\farcs5  to  3\arcsec)  calibration
binaries.   Their  motion  is  accurately modeled  based  on  previous
observations at SOAR.  The models are adjusted iteratively (the latest
adjustment in 2019 November).   Measurements of those wide calibrators
by Gaia \citep{Gaia} show very small systematic errors of these models
\citep{SAM18}.  Typical rms  deviations of the observations of calibrators
from their  models are 0\fdg2 in angle  and 1 to 3  mas in separation.
The astrometric  accuracy strongly depends on  the target characteristics
(larger errors at large $\Delta m$ and for faint pairs), as well as on
the seeing  and telescope vibration. The contrast  limit for companion
detection also depends on the  conditions, so that difficult pairs can
be resolved in one observing run and unresolved in another run.

\section{Results}
\label{sec:res}

\subsection{Data Tables}

The  results  (measures of  resolved  pairs  and non-resolutions)  are
presented in  exactly the same format  as in \citet{SAM19}.  The  long tables
are published electronically; here we describe their content.

\begin{deluxetable}{ l l  l l }
\tabletypesize{\scriptsize}
\tablewidth{0pt}
\tablecaption{Measurements of double stars at SOAR 
\label{tab:measures}}
\tablehead{
\colhead{Col.} &
\colhead{Label} &
\colhead{Format} &
\colhead{Description, units} 
}
\startdata
1 & WDS    & A10 & WDS code (J2000)  \\
2 & Discov.  & A16 & Discoverer code  \\
3 & Other  & A12 & Alternative name \\
4 & RA     & F8.4 & R.A. J2000 (deg) \\
5 & Dec    & F8.4 & Declination J2000 (deg) \\
6 & Epoch  & F9.4 & Julian year  (yr) \\
7 & Filt.  & A2 & Filter \\
8 & $N$    & I2 & Number of averaged cubes \\
9 & $\theta$ & F8.1 & Position angle (deg) \\
10 & $\rho \sigma_\theta$ & F5.1 & Tangential error (mas) \\
11 & $\rho$ & F8.4 & Separation (arcsec) \\
12 &  $\sigma_\rho$ & F5.1 & Radial error (mas) \\
13 &  $\Delta m$ & F7.1 & Magnitude difference (mag) \\
14 & Flag & A1 & Flag of magnitude difference\tablenotemark{a} \\
15 & (O$-$C)$_\theta$ & F8.1 & Residual in angle (deg) \\
16 & (O$-$C)$_\rho$ & F8.3 & Residual in separation (arcsec) \\
17  & Ref. & A8   & Orbit reference\tablenotemark{b} 
\enddata
\tablenotetext{a}{Flags: 
q -- the quadrant is determined; 
* -- $\Delta m$ and quadrant from average image; 
: -- noisy data or tentative measures. }
\tablenotetext{b}{References  are provided at
  \url{https://www.astro.gsu.edu/wds/orb6/wdsref.txt} }
\end{deluxetable}

Table~\ref{tab:measures}  lists 1735 measures  of 1288  resolved pairs
and subsystems,  including new discoveries.  The  pairs are identified
by their WDS-style codes based on the J2000 coordinates and discoverer
designations adopted  in the  WDS catalog \citep{WDS},  as well as by
alternative   names  in   column  (3),   mostly  from   the  Hipparcos
catalog.  Equatorial coordinates for  the epoch  J2000 in  degrees are
given  in  columns (4)  and  (5)  to  facilitate matching  with  other
catalogs and databases.  In the case of resolved multiple systems, the
position measurements  and their errors (columns  9--12) and magnitude
differences  (column  13) refer  to  the  individual pairings  between
components, not to their photo-centers.   As in the previous papers of
this  series, we  list  the  internal errors  derived  from the  power
spectrum model  and from the difference between  the measures obtained
from two data  cubes.  The real errors are  usually larger, especially
for  difficult pairs  with substantial  $\Delta m$  and/or  with small
separations.    Residuals  from   orbits  and   from  the   models  of
calibrators,  typically between  1  and 5  mas  rms, characterize  the
external errors of the HRcam astrometry.

The  flags in column  (14) indicate the cases where  the true  quadrant is
determined (otherwise the position angle is measured modulo 180\degr),
when the  relative photometry of wide  pairs is derived from  the long-exposure
images (this  reduces the bias  caused by speckle  anisoplanatism), and
when the data are noisy  or the resolutions are tentative (see TMH10).
For binary stars with known  orbits, the residuals to the latest orbit
and its reference are provided in columns (15)--(17). 

Non-resolutions  are  reported  in Table~\ref{tab:single}.  Its  first
columns  (1)  to   (8)  have  the  same  meaning   and  format  as  in
Table~\ref{tab:measures}.  Column  (9)  gives the  minimum  resolvable
separation when  pairs with $\Delta m  < 1$ mag are  detectable. It is
computed from  the maximum spatial  frequency of the useful  signal in
the power  spectrum and  is normally close  to the  formal diffraction
limit  $\lambda/D$. The following  columns (10)  and (11)  provide the
indicative  dynamic range,  i.e. the  maximum magnitude  difference at
separations  of 0\farcs15  and 1\arcsec,  respectively, at  $5\sigma$
detection level.  The last  column (12) marks  noisy data by  the flag
``:''.

\begin{deluxetable}{ l l  l l }
\tabletypesize{\scriptsize}
\tablewidth{0pt}
\tablecaption{Unresolved stars 
\label{tab:single}}
\tablehead{
\colhead{Col.} &
\colhead{Label} &
\colhead{Format} &
\colhead{Description, units} 
}
\startdata
1 & WDS    & A10 & WDS code (J2000)  \\
2 & Discov.  & A16 & Discoverer code  \\
3 & Other  & A12 & Alternative name \\
4 & RA     & F8.4 & R.A. J2000 (deg) \\
5 & Dec    & F8.4 & Declination J2000 (deg) \\
6 & Epoch  & F9.4 & Julian year  (yr) \\
7 & Filt.  & A2 & Filter \\
8 & $N$    & I2 & Number of averaged cubes \\
9 & $\rho_{\rm min}$ & F7.3 & Angular resolution (arcsec)  \\
10&  $\Delta m$(0.15) & F7.2 & Max. $\Delta m$ at 0\farcs15 (mag) \\
11 &  $\Delta m$(1) & F7.2 & Max. $\Delta m$ at 1\arcsec (mag) \\
12 & Flag & A1 & : marks noisy data  
\enddata
\end{deluxetable}

Table~\ref{tab:measures}  contains  162 pairs  resolved  for the  first
time;  some of  those  were confirmed  in  subsequent observing  runs.
In the following sub-sections we discuss the new pairs.

\subsection{Young Moving Groups and Associations}
\label{sec:YMG}


\startlongtable  
\begin{deluxetable}{ l l c  c  l }
\tabletypesize{\scriptsize}
\tablewidth{0pt}
\tablecaption{New YMG Pairs
\label{tab:YMG}}
\tablehead{
\colhead{WDS} &
\colhead{TIC} &
\colhead{$\rho$} &
\colhead{$\Delta I$} &
\colhead{$N$\tablenotemark{a}} \\
 &     &    
 \colhead{(arcsec)} &
 \colhead{(mag)} & 
}
\startdata
00376$-$2709  &   246895416  &  0.14 &2.5 & 4 \\
02002$-$8025  &   273789976  &  0.03 &0.0 & 3 \\
02109$-$4604  &   7242537    &  0.21 &2.9 & 2 \\
02489$-$3404  &   122671519  &  0.05 &0.0 & 2 \\ 
02568$-$6343  &   220556639  &  0.07 &0.0 & 3 \\
03165$-$3541  &   176832633  &  0.19 &4.4 & 3 \\
03259$-$3556  &   142874733  &  0.06 &2.3 & 3 \\
04001$-$2902  &   44670258   &  0.07 &2.4 & 1? \\ 
04084$-$2745  &   44793998   &  0.37 &1.5 & 2 \\
04316$-$3043  &   170699229  &  0.44 &1.6 & 1 \\
04536$-$2836  &   671393     &  0.13 &0.0 & 2 \\
05085$-$2102  &   146539195  &  0.05 &0.6 & 3 \\
05287$-$3327  &   24448282   &  0.14 &1.2 & 2 \\ 
05371$-$3932  &   144499196  &  0.16 &1.3 & 2 \\
05412$-$4118  &   21438160   &  0.07 &0.0 & 3 \\ 
05425$-$1535  &   46739994   &  1.15 &0.7 & 1 \\
05471$-$3211  &   100608178  &  0.27 &1.0 & 1 \\
05473$-$5450  &   350563576  &  0.84 &5.0 & 1 \\
05504$-$2915  &   32930236   &  0.85 &3.6 & 1 \\
05597$-$6209  &   149935360  &  1.15 &4.0 & 1 \\
06086$-$3403  &   201391310  &  0.57 &3.1 & 1 \\
06086$-$5704  &   260127241  &  0.12 &0.0 & 1 \\
06220$-$7932  &   270424741  &  0.06 &0.0 & 1 \\
06462$-$8359  &   397231463  &  1.45 &5.7 & 1 \\
07019$-$3922  &   157212164  &  1.83 &4.1 & 1 \\
07147$-$4010  &   22766740   &  0.16 &0.1 & 2 \\
07310$-$8419  &   405077613  &  0.61 &3.8 & 1 \\
07336$-$4019  &   173957127  &  1.70 & $-$0.15 & 1* \\  
07336$-$4019  &   173957127  &  0.12 &0.6 & 1* \\  
07406$-$6704  &   300741570  &  2.27 &3.4 & 1 \\
07418$-$4630  &   123642034  &  1.49 &1.2 & 1 \\
07437$-$6107  &   281582156  &  0.18 &3.4 & 2 \\
07571$-$2227  &   142844055  &  1.16 &1.1 & 1 \\
08262$-$3902  &   183974196  &  0.24 &0.1 & 1 \\
09095$-$5538  &   385012516  &  0.05 &0.1 & 4 \\
10054$-$7137  &   372515598  &  0.10 &0.0 & 2 \\ 
10056$-$5731  &   462492721  &  3.85 &4.0 & 1 \\  
10074$-$4622  &   311258541  &  0.18 &0.1 & 2 \\
10207$-$6311  &   378126824  &  0.32 &1.4 & 2 \\
10330$-$6144  &   460604193  &  0.52 &4.5 & 1 \\ 
11081$-$6342  &   466799461  &  2.37 &5.4 & 1 \\ 
11098$-$3828  &   151738485  &  0.05 &1.1 & 1? \\ 
11099$-$3739  &   151762498  &  0.06 &0.6 & 1 \\
11262$-$5823  &   451452509  &  0.11 &1.6 & 1 \\
11545$-$5325  &   400913139  &  0.03 &0.7 & 1? \\ 
12172$-$1033  &   203233128  &  0.12 &0.4 & 1 \\
12269$-$3316  &   130722957  &  0.67 &0.1 & 1 \\
12328$-$7654  &   360339486  &  0.17 &1.0 & 2 \\
12559$-$7417  &   361525866  &  1.10 &0.7 & 2 \\
12577$-$6652  &   335287811  &  0.13 &0.3 & 3 \\ 
13090$-$5720  &   253501247  &  4.10 &0.9 & 1 \\ 
13123$-$5441  &   406376573  &  1.00 &0.9 & 1 \\
13137$-$5807  &   406694754  &  0.68 &1.0 & 2 \\
13260$-$5112  &   438733975  &  0.11 &1.1 & 3 \\
13271$-$4856  &   438790187  &  0.95 &2.5 & 2 \\
13275$-$4719  &   438627800  &  2.82 &4.4 & 1 \\
13341$-$5624  &   457308993  &  1.12 &0.1 & 1* \\      
13341$-$5624  &   457308993  &  0.09 &0.0 & 1* \\  
13413$-$4537  &   243415454  &  3.59 &1.6 & 1 \\
13415$-$4431  &   243425206  &  2.47 &0.0 & 1* \\    
13415$-$4431  &   243425206  &  0.66 &0.1 & 1* \\    
13453$-$4102  &   166302995  &  0.31 &3.1 & 1 \\
13485$-$6727  &   429383724  &  0.71 &2.8 & 1 \\
13491$-$4413  &   243621789  &  2.92 &4.9 & 1 \\
13523$-$3826  &   166624597  &  0.13 &1.3 & 1 \\
13538$-$5502  &   208387087  &  3.17 &0.5 & 1 \\
13579$-$4432  &   359830202  &  0.17 &2.2 & 1 \\
14028$-$1850  &   6119516    &  1.80 &0.8 & 1 \\
14161$-$4031  &   179793360  &  0.17 &2.2 & 2 \\
14169$-$3648  &   179819049  &  0.08 &0.0 & 1 \\
14171$-$4038  &   179829109  &  0.96 &3.2 & 1 \\
14241$-$3923  &   167542104  &  1.50 &1.6 & 1 \\
14381$-$4322  &   128453434  &  1.11 &1.6 & 1 \\
14463$-$5056  &   250091359  &  0.63 &2.2 & 1 \\
14535$-$3903  &   160451137  &  1.04 &2.6 & 1 \\
14541$-$3606  &   160574439  &  0.08 &0.8 & 1 \\
14544$-$3718  &   160576551  &  3.26 &0.6 & 1 \\
14592$-$4012  &   121196256  &  2.50 &0.9 & 1 \\
15180$-$3335  &   272248916  &  0.08 &0.0 & 2 \\ 
15206$-$3132  &   460325085  &  1.22 &2.1 & 1 \\
15230$-$3052  &   54077774   &  2.17 &0.6 & 1* \\   
15230$-$3052  &   54077774   &  0.36 &1.1 & 1* \\    
15233$-$3127  &   54077130   &  3.35 &0.2 & 1 \\
15280$-$3208  &   54512674   &  2.22 &2.4 & 1 \\
15299$-$3136  &   54667962   &  1.73 &2.4 & 1 \\
15312$-$3505  &   54802536   &  0.08 &0.7 & 2 \\
15476$-$3127  &   442571495  &  0.07 &0.0 & 1 \\
16186$-$3839  &   318141352  &  1.66 &5.2 & 1 \\
16210$-$0617  &   135890809  &  0.05 &0.2 & 1 \\
16223$-$3843  &   4061225    &  0.81 &1.9 & 1 \\
16338$-$5119  &   22836043   &  1.81 &3.1 &  1* \\  
16338$-$5119  &   22836043   &  0.13 &1.1 & 1* \\    
16345$-$1106  &   152667565  &  1.95 &0.7 & 1 \\
16361$-$1324  &   414338264  &  1.21 &5.6 & 1 \\
16498$-$1239  &   398869084  &  0.60 &3.8 & 1 \\
16502$-$1108  &   181292505  &  0.76 &2.4 & 1 \\
17076$-$0515  &   142638811  &  0.23 &2.5 & 1 \\
17123$-$1131  &   146003265  &  0.09 &0.0 & 1 \\
17142$-$0038  &   176322832  &  0.48 &2.9 & 1 \\
17185$-$7858  &   384747990  &  0.12 &0.4 & 1 \\
17563$-$5833  &   337276808  &  1.85 &1.3 & 1 \\
20108$-$3845  &   269768590  &  1.37 &2.6 & 2 \\
20146$-$5431  &   201751726  &  0.58 &0.0 & 1 \\
21123$-$8129  &   403995704  &  0.10 &0.0 & 1 \\ 
21210$-$5229  &   79403459   &  0.09 &0.2 & 4 \\  
21215$-$6655  &   419610508  &  0.06 &1.3 & 1 \\ 
21589$-$4706  &   389726031  &  0.15 &0.0 & 3   
\enddata 
\tablenotetext{a}{The question mark indicates unreliable resolutions, asterisks distinguish
 triple stars.}
\end{deluxetable}

\begin{figure}
\epsscale{1.1}
\plotone{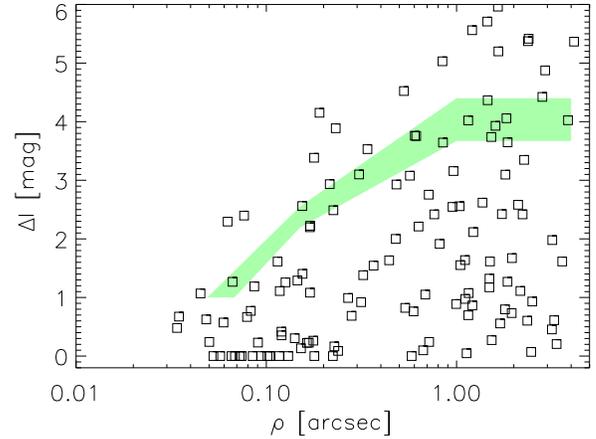}
\caption{Resolved pairs in the YMG sample: $\Delta I$
  vs. separation. The shaded green area depicts the lower and upper
  quartiles of the detection limits.
\label{fig:YMGplot} }
\end{figure}

The TESS follow-up  program started in 2018. It  was complemented by a
sample of the  members of young moving groups  and associations (YMGs)
based  on the  TESS input  catalog, TIC  \citep{Stassun2019}  and Gaia
astrometry.  These objects are identified here by TIC numbers preceded
by  the  letter `T'.  Overall,  892  objects  from this  program  were
observed at least once. The  largest number of observations (608) were
secured in 2019, and only 32 in 2020; the total number of observations
is 985 (some newly resolved pairs were re-visited).

Figure~\ref{fig:YMGplot} plots the magnitude difference $\Delta I$
vs. separation $\rho$ for resolved pairs of the YMG sample. One
notes the paucity of pairs with separations from 0\farcs2 to 0\farcs7
and a small $\Delta I$. Such near-equal pairs do not have astrometry in
Gaia and, for this reason, were not included in the sample which,
therefore, is biased against such binaries. 

There are 129 resolved pairs  among 892 YMG objects, so the observed raw
multiplicity rate is 14.5 per cent.  Most of these pairs (103) are new
discoveries.  Statistical analysis of  the multiplicity is outside the
scope  of  this paper,  which  only  reports  the observations.   Some
companions, especially those with large  $\rho$ and $\Delta I$, can be
unrelated  stars (optical  pairs). All  newly resolved  YMG  pairs are
listed in  Table~\ref{tab:YMG}. They  are identified by  the WDS-style
code  and  the  TIC   numbers.   The  following  columns  contain  the
separation $\rho$, the magnitude difference $\Delta I$, and the number
of visits $N$ where the pair was resolved. Mostly, we re-visited close
pairs and  found that some  of them show  rapid orbital motion  on the
time  span of  1--3 yrs.  They are  promising candidates  for future
orbit determinations and measurements of masses.  Three new close pairs were
not resolved  in subsequent visits  either because they  moved under
the resolution limit or  because the first resolutions were  
unreliable.  These stars  can be in fact single; they  are distinguished by 
question marks in the last column, and further observations are needed
to confirm these pairs.

The resolution  and contrast  limits depend on  the seeing  and target
magnitude. One might think that  brighter targets have a larger chance
of  binary  detection. However,  the  median  TESS  magnitudes $T$ of  the
resolved and  unresolved targets are 11.4 and  11.5 mag, respectively,
and their  distributions look alike. Therefore, the  magnitude bias is
small, if any.   The total range of magnitudes in  this sample is from
$T=6$ to $T =13$ mag.

\begin{figure}
\epsscale{1.1}
\plotone{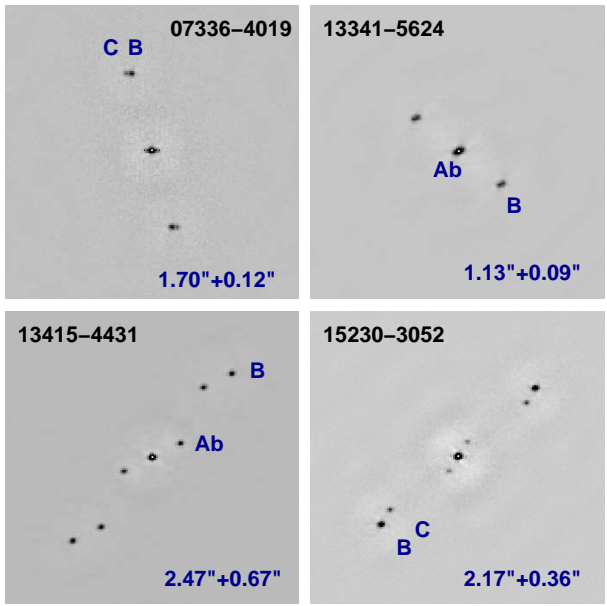}
\caption{Four new triples  in the YMG sample. The  panels show speckle
  ACFs  (in  negative rendering)  in  the  full  3\farcs15 field.  The
  letters  mark secondary  peaks corresponding  to the  companions (as
  opposed to the mirror peaks) inferred from the SAA images.   The
    separations of the  outer and inner pairs in  arcseconds are given
    in the lower-right corner. 
\label{fig:YMGtriples} }
\end{figure}

Five YMG targets  turned out  to be resolved  triple stars. They  have two
entries  in   Table~\ref{tab:YMG}  (one  per   subsystem),  marked  by
asterisks  in the last  column. Figure~\ref{fig:YMGtriples}  illustrates the
ACFs of four new young triples out of five.

\subsection{Other New Pairs}
\label{sec:new}

\begin{deluxetable}{ l l c  cc l  }
\tabletypesize{\scriptsize}
\tablewidth{0pt}
\tablecaption{New Double Stars
\label{tab:binaries}}
\tablehead{
\colhead{WDS} &
\colhead{Name} &
\colhead{$\rho$} &
\colhead{$\Delta m$} &
\colhead{$N$} &
\colhead{Prog.\tablenotemark{a}} \\
 &     &    
 \colhead{(arcsec)} &
 \colhead{(mag)} & 
}
\startdata
00111$+$0513  &   HIP 898     &0.12 &2.5& 2 & NKD \\ 
01027$-$2519  &   HIP 4874    &0.10 &3.4& 6 & NKD \\ 
01262$+$1349  &  HIP 6705     &0.79 &3.6& 2 & NKD \\  
01406$+$0846  &  HIP 7819     &0.09 &1.3& 2 & REF \\
02035$-$0455  &   HIP 9603    &0.20 &2.9& 4 & NKD \\ 
02324$+$0323  & HIP 11815     &1.50 &3.1& 2 & NKD \\
03225$+$1744  &   HIP 15724   &1.70 &7.4& 2 & NKD \\ 
04007$-$2305  &   GJ 3260AB   &1.35 &2.6& 7 & Pan \\ 
04007$-$2305  &   GJ 3260BC   &0.33 &0.1& 7 & Pan \\ 
04141$-$3155  & TIC168789840  &0.42 &0.3& 4 & TESS \\
04234$+$1546  &  HIP 20485    &1.20 &4.5& 2 & NKD \\
04279$+$2427  &  HIP 20834    &0.17 &2.7& 4 & NKD \\
04330$-$1633  &   HIP 21222  &0.09 &1.7& 7 & NKD \\ 
04518$+$1339  &   BU 552Ba,Bb &0.04 &1.2& 2 & MSC \\ 
06006$-$5806  & HIP 28464     &0.44 &4.5 & 2 & REF \\ 
06215$+$1718  &   HIP 30220   &0.92 &3.0& 3 & NKD \\ 
06443$-$2349  &   TDS4085BC   &0.33 &0.4& 1 & TESS \\ 
07151$+$1556  &   HIP 35071   &1.10 &2.6& 2 & NKD \\  
07390$+$1913  &   HIP 37246   &0.44 &1.4& 3 & NKD \\ 
07584$-$1501  &   HIP 38969   &0.11 &3.1& 2 & NKD \\ 
08155$+$0959  &   HIP 40449   &0.23 &0.1& 3 & NKD \\ 
08187$-$1512  &   HIP 40724   &0.07 &2.3& 8 & NKD \\ 
08253$+$0415  &   HIP 41277   &0.35 &3.7& 3 & NKD \\ 
08430$+$2408  &   TOK 265Aa,Ab   &0.18 &2.2& 4 & NKD \\ 
09095$-$0024  &   HIP 44953   &1.38 &3.7& 2 & NKD \\  
09308$+$1815  &   HIP 46662   &0.14 &2.7& 5 & NKD \\ 
09361$-$5145  &  RST 415Aa,Ab &0.14 &1.7& 1 & TESS \\ 
09380$+$2231  &   HIP 47261   &1.11 &3.7& 2 & NKD \\  
09429$-$5502  &  RST3660Aa,Ab &0.07 &1.3& 3 & WDS \\ 
09527$-$7933  &   KOH 86Aa,Ab &0.04 &1.7& 1 & MSC  \\ 
10041$+$1848  &   HIP 49324   &0.13 &0.1& 6 & NKD  \\ 
10211$-$1744  &   HIP 50696   &0.59 &1.2& 3 & NKD \\ 
10212$-$5143  &   I 853Ba,Bb  &0.09 &1.0& 3 & WDS \\   
10262$-$6318  &   HIP 51083   &0.10 &2.5& 2 & REF \\
10283$-$2416  &   HIP 51263   &0.20 &2.0& 6 & NKD  \\ 
10527$+$0029  &   HIP 53175AB &1.65 &3.0& 4 & NKD \\ 
10527$+$0029  &   HIP 53175BC &0.17 &1.2& 4 & NKD  \\ 
11100$-$1017  &   HIP 54569AB &0.37 &3.1& 4 & NKD \\
11100$-$1017  &   HIP 54569Aa,Ab &0.04 &0.1& 4 & NKD \\
11358$+$2437  &   HIP 56570   &0.39 &3.4& 2 & NKD \\ 
11418$+$0508  &   HIP 57058   &0.06 &0.0& 7 & NKD \\ 
11563$+$1102  &   SKF 256Aa,Ab   &0.89 &4.2& 2 & NKD \\
12104$-$4352  &   HD 105750   &0.06 &1.3&13 & Pan \\ 
12356$-$3454  &   GJ 1161B    &0.33 &0.1& 7 & Pan  \\ 
13344$-$2730  &   HIP 66229   &0.04 &1.5& 3 & NKD \\ 
14106$-$2826  &   HIP 69249   &0.24 &2.6& 4 & NKD \\ 
14232$-$6302  & FIN 221Aa,Ab  &0.10 &2.2& 1 & TESS \\ 
15003$+$0739  &   HIP 73424   &0.18 &4.6& 1?& REF \\ 
15481$-$5811  & SKF2839Aa,Ab    &0.26 &1.5& 5 & Pan \\  
16238$-$0258  &  HIP 80315    &0.20 &3.8& 2 & NKD \\
17190$-$4638  &   HD 156274B  &0.04 &1.9& 8 & Pan \\ 
17331$-$3035  &   CHM 6BC     &3.09 &5.0& 2 & NKD \\ 
18185$-$3441  &   HIP 89708   &0.18 &3.5& 1 & REF \\  
19443$-$2657  &   HD 186265   &1.06 &5.0& 5 & Pan  \\ 
20118$-$3825  &   RTW2011AB   &4.23 &0.1& 1 & NKD \\ 
22412$-$1625  &   RTW2241AB   &4.48 &0.1& 1 & NKD \\
22590$-$0432  &   BD$-$05 5901  &0.42 &1.2& 1 & MSC \\ 
23231$-$7747  &   UC 4934Aa,Ab  &0.12 &0.4& 1 & HIP  \\ 
23343$+$0932  &   HIP 116334  &0.37 &3.7& 4 & NKD   
\enddata 
\tablenotetext{a}{
HIP -- {\it Hipparcos} suspected binary;
NKD -- nearby K-dwarfs;
MSC -- multiple system; 
Pan -- program by B.~Pantoja;
REF-- reference star;
TESS -- TESS follow-up;
WDS -- neglected pair.
}
\end{deluxetable}

\begin{figure*}
\epsscale{0.8}
\plotone{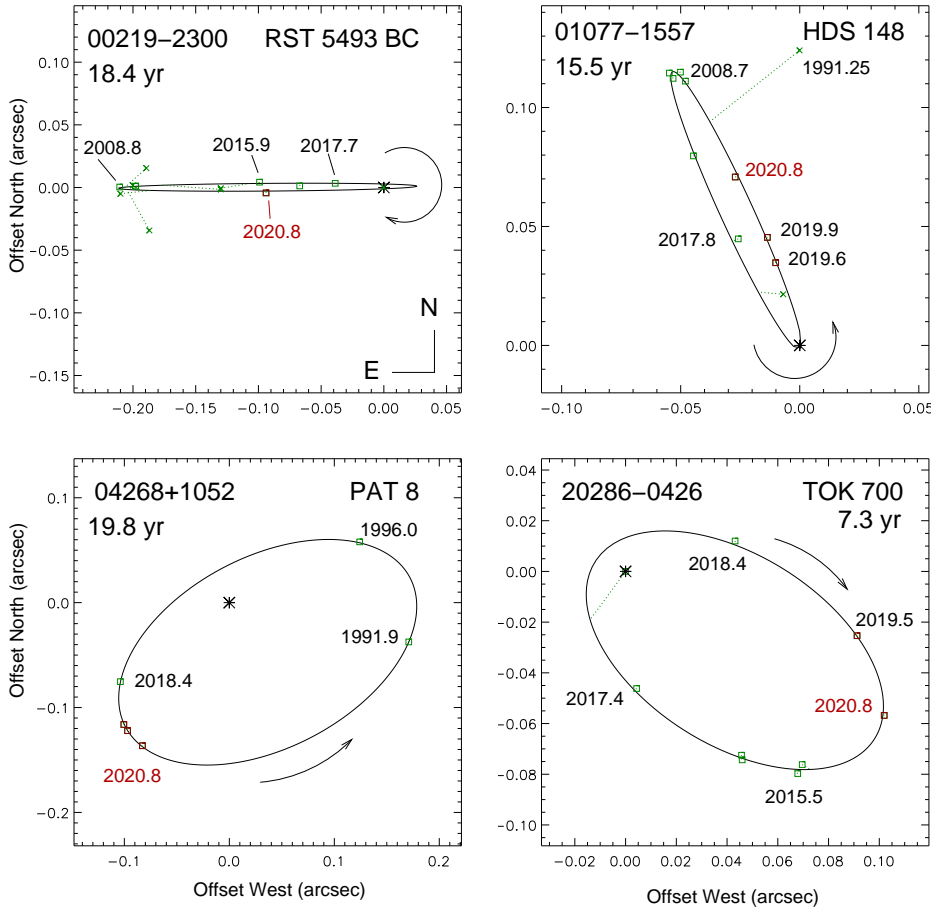}
\caption{Four first-time orbits computed using the SOAR observations
  in 2020.  Accurate speckle measurement are plotted as squares (in
  red after 2019.0), visual micrometer measurements as crosses. The
  axis scale is in arcseconds. 
\label{fig:orb} }
\end{figure*}
\vspace*{1cm}

Table~\ref{tab:binaries}  highlights  59  pairs  resolved in  2020  or
resolved  earlier but not  yet published.   All measurements  of these
pairs are found in Table~\ref{tab:measures}.  Table~\ref{tab:binaries}
is similar  to Table~\ref{tab:YMG}, but it  contains an additional column
specifying the  program.  The largest  number of new pairs,  36, comes
from the  survey of  K-type dwarfs. Most  of these were  observed more
than once,  and some (the closest)  are in rapid  orbital motion.  The
combined spectro-interferometric orbit of one such pair, HIP~57058, is
determined  (Figure~\ref{fig:57058}).   Five  pairs are  serendipitous
resolutions of reference  stars (three of those predate  2020 but were
not reported  previously), two are  new close subsystems  in classical
visual binaries  (WDS program), and  three are subsystems  in multiple
stars  (MSC). Most new  pairs are  real physical  binaries, and  a few
appear to be optical (chance projections).

\subsection{New and updated orbits}
\label{sec:orbits}

\begin{figure}
\epsscale{1.1}
\plotone{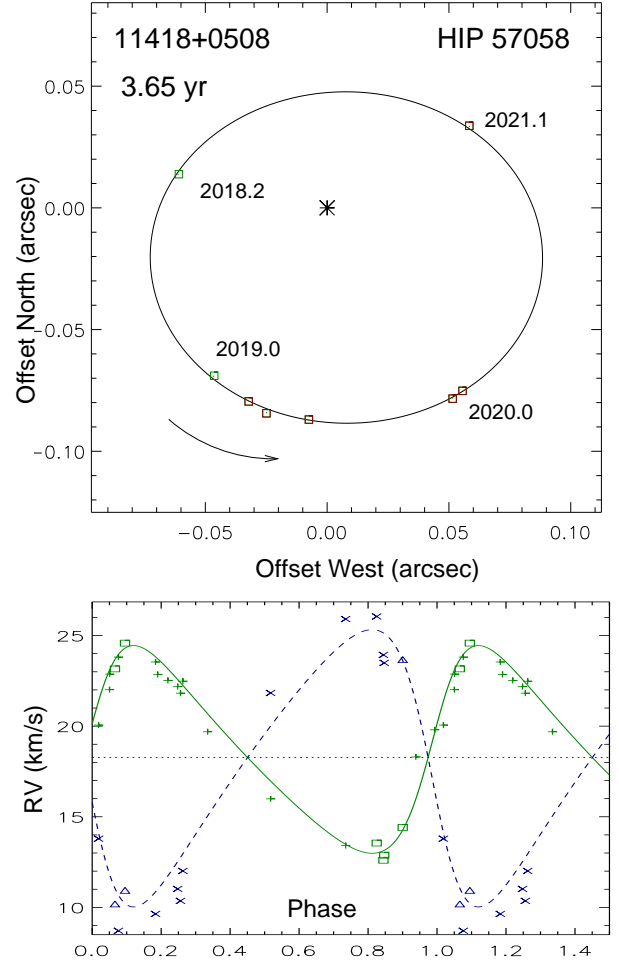}
\caption{Orbit of HIP 57058 in the plane of the sky (top) and its RV
  curve (bottom).
\label{fig:57058} }
\end{figure}

New  positional measurements  furnished  by the  SOAR speckle  program
provide material for calculation  of new visual orbits and improvement
of the  known ones.  The  previous paper of this  series \citep{SAM19}
gave a long  list of new orbits.  Here we only  give references on the
latest orbits  resulting from this  program \citep{Mendez2021, TL2020,
  Tok2021a, Circ203} and provide examples in Figures~\ref{fig:orb} and
\ref{fig:57058}.  The  orbital elements  are  published,  so
there is no need to repeat them here. We comment on each pair below.

{\it 00219$-$2300}  (ADS 302, HIP  1732) is a  triple system at  60 pc
from the Sun. The pair BC is  located at 6\farcs1 from the main star A
and composed of similar  K-type dwarfs. Five micrometric measures made
since its discovery by S.~Rossiter  in 1949 were insufficient for
orbit calculation,  and we see  why: the orbit  is oriented edge-on and  has a
large  eccentricity $e=0.83$. The  pair was  monitored at  SOAR since  2008; it
closed  down and  opened again  this  year after  passing through  the
periastron. The mass sum is 1.5 \msun.

{\it 01077$-$1557} (HIP 5295) is a pair of solar-type dwarfs
discovered by Hipparcos. Its first 15.5  yr orbit is based exclusively
on the 10 SOAR measures because the Hipparcos measurement appears to
be misleadingly inaccurate. We observed the pair in 2008 near maximum
separation; it passed through the periastron of eccentric ($e=0.98$) orbit in
2019.1 (an uncertain measure was attempted in 2018 below the diffraction limit)
and became resolved again in 2019. Although this is a first-time
orbit, its elements are quite accurate.   

{\it 04268+1052} (HIP 20751) is a K0V binary dwarf in the Hyades. Only
two   measurements  were   available  before   2018,  when   the  SOAR
observations started as  part of the K-dwarf survey.  The arc observed
at SOAR is quite short,  but, combined with the historic measurements,
it allows the calculation of the  first 20 yr orbit. The periastron in
2017.4  was, unfortunately,  missed, and  now the  pair is  on  a slow
segment of its orbit. The fit  of the 7 orbital elements to the 6 position
measurements is nearly perfect (residuals less than 1 mas).

{\it 20286$-$0426} (HIP 100988) is another pair of solar-type stars
discovered at SOAR in 2015.5. Its monitoring to date allows
calculation of the first 7.3 yr orbit, now almost completely covered. 

These four  pairs chosen  to illustrate new  orbits have  something in
common.  They  are composed  of low-mass stars  and their  orbits have
substantial eccentricities,  from 0.65  to 0.98.  Radial velocity (RV)
monitoring near the periastron  can furnish direct measurements of the
mass ratios and orbital  parallaxes.  Such observations can be planned
in the  future, knowing the visual  elements. One notes that  the  11 yr
duration of the extended  Gaia mission  is not  long enough  to derive
astrometric orbits from the photo-center motion, while these pairs are
too  close   for  a  direct  resolution  by   Gaia.   However,  future
combination of speckle orbits  and Gaia astrometry will allow accurate
modeling of the photo-center  motion, leading to unbiased measurements
of the parallaxes and, hence, masses.

Figure~\ref{fig:57058} illustrates the synergy between interferometric
and spectroscopic  data for the case  of a nearby  K4V dwarf HIP~57058
(GJ  435.1,  distance 31\,pc).   It  was  previously  identified as  a
double-lined spectroscopic  binary. A preliminary  spectroscopic orbit
with   a  period   of  725.9   days   (2.00  yr)   was  published   by
\citet{Sperauskas2019}, who also mention  its first resolution at SOAR
in 2018.2.  The  pair was also resolved in 2016  by T.~Henry, but this
measurement  is  not  yet  published.  Continued  monitoring  at  SOAR
revealed  that the observed  motion is  not compatible  with the  2 yr
period, and  the true period  is two times  longer. Stars A and  B are
similar  ($\Delta I =  0.3$ mag),  and a  wrong attribution  of radial
velocities  (RVs) to  a particular  component was  the reason  for the
incorrect  spectroscopic  orbit  (which  is single-lined  despite  the
double-lined nature  of the system).   The new orbit with  $P=3.65$ yr
presented in  Figure~\ref{fig:57058} uses only RVs  of both components
when they were resolved  spectroscopically.  The RV curves look noisy,
but,  considering the  small amplitudes  (5.7  and 7.7  \kms) and  the
difficulty  of measuring  blended spectra,  barely resolved  only near
the RV maximum, the rms residuals  of 0.4 and 1.2 \kms appear quite
acceptable.  The residuals of  the interferometric measures are small,
under 1\,mas.  The  mass sum of 1.35 \msun computed  from the orbit and
the Gaia parallax matches the RV amplitudes and the spectral type.

\subsection{Hierarchical systems}
\label{sec:mult}

Accumulation of accurate speckle measures with dense coverage made at
SOAR allows the study of relative motions in hierarchical stellar
systems. The latest papers in this area were already cited
\citep{TL2020,Tok2020,Tok2021a}. Here two additional examples are given.

\begin{figure}
\epsscale{1.1}
\plotone{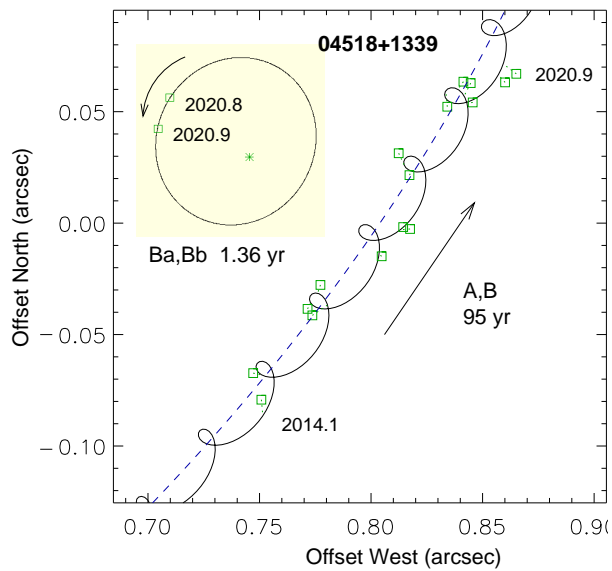}
\caption{Wavy motion of the visual binary J04518+1339 (BU 552) caused
  by the subsystem Ba,Bb, and the orbit of Ba,Bb (insert) deduced from
  the RVs and two measures at SOAR.
\label{fig:BU552} }
\end{figure}

{\it  J04518+1339}  (HD  30869,  HIP  22607)  is  a  quadruple  system
belonging  to the  Hyades cluster.   The outer  visual pair  BU~552 has been
known  since  1877, and  its  95  yr orbit  is  very  well defined  by
observations  covering  1.5  revolutions.   Components  A  and  B  are
double-lined spectroscopic  binaries with  periods of 143.6  and 496.7
days, respectively \citep{Tomkin2007}. The estimated semimajor axis of
Ba,Bb  is  33\,mas,  favoring  detection  of  wobble  caused  by  this
subsystem, while the  astrometric orbit of Aa,Ab with  an amplitude of
4.4\,mas was  computed by \citet{Ren2013}.  From 2016  on, this binary
was frequently observed at SOAR. In 2020, the slight elongation of the
secondary  ACF peak  was noted  and  a triple-star  model was  fitted,
tentatively resolving Ba,Bb on  two occasions.  Similar elongation can
be  suspected upon  examination of  previous observations,  but  it is
often  concealed by  false elongation  due to  telescope  vibration or
charge-transfer problem;  only observations of the  highest quality in
the filter $y$  allow marginal resolutions of Ba,Bb  at phases near its
maximum separation.  Figure~\ref{fig:BU552}  shows the two measures of
Ba,Bb that fit nicely  the orbit of \citet{Tomkin2007} with additional
elements $a= 34$\,mas, $\Omega  = 326\degr$, and $i=32\degr$. The axis
and inclination  match their estimates derived  from the spectroscopic
orbit. Moreover, the wobble in the motion of A,B with a period of 1.36
yr  is  rather obvious;  its  amplitude  is  9.7\,mas. Motion  of  the
photocenter of A with a 143 day period increases the residuals.

\citet{Tomkin2007} estimate orbital inclination  of Aa,Ab to be around
49\degr,  suggesting  possible  coplanarity  with  the  orbit  of  A,B
(inclination 51\degr). \citet{Ren2013} found that the nodes of A,B and
Aa,Ab  have  similar position  angles,  but  they  give a  mismatching
inclination of 94\fdg4 for Aa,Ab. Our work establishes the orientation
of  the  orbit of  Ba,Bb:  it  is inclined  to  the  orbit  of A,B  by
26\degr.  These  preliminary  results  should be  refined  by  further
observations, preferably with larger aperture (the resolution of Ba,Bb
at SOAR is just marginal).

\begin{figure}
\epsscale{1.1}
\plotone{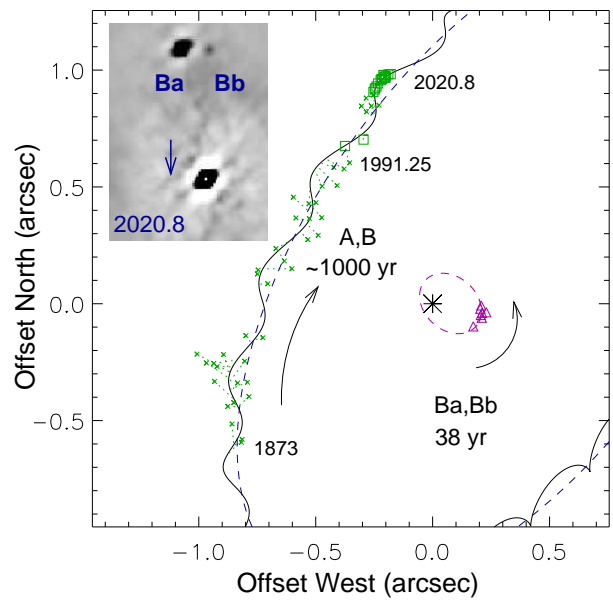}
\caption{Orbital motion of J02460$-$0457.  The wavy line is the motion
  of A,B with wobble.  Squares depict accurate measures from Hipparcos
  and speckle, crosses are micrometer measures.  The inner orbit Ba,Bb
  is plotted  around the center on  the same scale  by magenta ellipse
  and triangles. The  insert shows the speckle ACF  recorded in 2020.8
  where the blue arrow shows the missing ACF peak (see text).
\label{fig:BU83} }
\end{figure}

{\it J02460$-$0457} (HD 17251, HIP 12912) is a triple system where the
outer pair  A,B (BU~83) has been known  since 1873. A  third faint component
was  discovered  at  SOAR  in  2016  and  attributed  to  the  primary
\citep{SAM17}, based on the sign of the wobble. Continued observations
show that this is not correct  and the faint companion in fact belongs
to   the  secondary  star   B.   The   situation  is   illustrated  in
Figure~\ref{fig:BU83}.   Generally, the ACF  of a  triple star  contains 6
secondary  peaks corresponding  to 6  vectors between its components. The
relative intensity of the peaks is proportional to the products of the
relative fluxes; the two weakest peaks that correspond to the pair of the faintest
stars often  are lost in  the noise, as  is the case here.   The arrow
shows  the position  of the  missing peak  corresponding to  the faint
companion. If  it were associated with  A, this missing  peak would be
stronger  than  the  well-visible   peak  Bb.  The  missing  peak  was
marginally  seen in the  discovery ACF,  prompting the  original wrong
attribution of the new companion.

The rotation  direction of Ba,Bb  pair is opposite to  A,B. Existing
data  were  reprocessed with  the  assumption  that  the companion  is
associated  with B,  resulting in  more accurate  measurements  of the
Ba,Bb positions.  The wave in  the A,B motion  does not depend  on the
choice, so the  original argument associating the companion  with A was
not valid.   If the new companion  were associated with  A, the wobble
should produce  a proper motion  (PM) anomaly (difference  between the
Gaia short-term PM and the  long-term PM of the photocenter) of about
2  \masyr, oriented  in the  declination direction.   The  measured PM
anomaly in declination is  0.15$\pm$0.14 \masyr, compatible with zero;
it proves that the subsystem is associated with component B.

The inner and outer orbits fitted to the available measures, shown
in Figure~\ref{fig:BU83}, are still preliminary. The opposite sense of
rotation excludes their coplanarity: the mutual inclination is either
130\degr ~or 95\degr. Unequal masses and misaligned, eccentric orbits
are signs of dynamical interactions that probably defined the
architecture of this triple system.


\section{Summary}
\label{sec:sum}

The  total number of  observations made  with HRCam  to date  is about
25,000.  This paper  documents the observations made in  2020, as well
as  earlier unpublished data.  The HRCam  at SOAR  is used  by various
programs, executed  in a concerted and optimized  way and complemented
by  the uniform  data reduction  and calibration  procedures.  Focused
initially on the determination  of visual orbits, the programs expanded
into surveys of binarity in various populations.  Still, the orbits of
both previously known pairs and  those discovered at SOAR remain
the major use of the HRCam data.

This paper presents results of  the large survey of YMG population and
of  the  K-type  dwarfs  in  the solar  neighborhood.   Both  programs
discovered  a substantial  number  of tight  pairs  with fast  orbital
motion. Their continued monitoring  will lead to orbit calculation and
measurements  of masses  in the  near future.  For example,  the first
orbit of the K-dwarf HIP~57058  presented above uses two years of SOAR
data in combination with the longer RV coverage. Several orbits of new
M-dwarf pairs with  short periods will be published  by E.~Vrijmoet et
al.  (2021, in preparation).

Flexibility  of  the   HRCam  observing  procedure  brings  unexpected
benefits.  When  the unusual  sextuple eclipsing system  TIC 168789840
(J04141$-$3155) was discovered, it could be quickly tested at SOAR and
resolved  into a  0\farcs4 pair  --- a  key piece  of  information for
unveiling     the    architecture     of     this    unique     object
\citep{Powell2021}.  Thus,  the  ongoing  SOAR speckle  program  is  a
backbone for testing future discoveries in a quick and efficient way.


\begin{acknowledgments} 

We thank the SOAR operators for efficient support of this program, and
the SOAR  director J.~Elias for allocating some  technical time.  This
work is based in part  on observations carried out under CNTAC programs
CN2019A-2, CN2019B-13, CN2020A-19, and CN2020B-10.

R.A.M.  and E.C.  acknowledge  support from the FONDECYT/CONICYT grant
\# 1190038.  T.~Henry appreciates support of this work through NSF
  grants AST-1517413 and AST-1910130. 

This work  used the  SIMBAD service operated  by Centre  des Donn\'ees
Stellaires  (Strasbourg, France),  bibliographic  references from  the
Astrophysics Data  System maintained  by SAO/NASA, and  the Washington
Double Star  Catalog maintained  at USNO.  This  work has made  use of
data   from   the   European   Space   Agency   (ESA)   mission   Gaia
(\url{https://www.cosmos.esa.int/gaia})  processed  by  the  Gaia  Data
Processing      and     Analysis      Consortium      (DPAC,     {\url
  https://www.cosmos.esa.int/web/gaia/dpac/consortium}). Funding for the
DPAC  has been provided  by national  institutions, in  particular the
institutions participating in the Gaia Multilateral Agreement.

\end{acknowledgments} 

{\it Facilities:}  \facility{SOAR}.



\end{document}